\documentclass[apj]{emulateapj}
\usepackage{graphicx}
\usepackage{color}
\usepackage[titletoc,toc,title]{appendix}

\shorttitle{Galactic discs mass transport}
\shortauthors{Prieto et al.}

\begin{document}

\title{How AGN and SNe feedback affect mass transport and black hole growth in high redshift galaxies}


\author{Joaquin Prieto and Andr\'{e}s Escala}
\affil{Departamento de Astronom\'{i}a, Universidad de Chile, Casilla 36-D, Santiago, Chile.}

\author{Marta Volonteri and Yohan Dubois}
\affil{CNRS and UPMC Universit\'e Paris 06, UMR 7095, Institut d'Astrophysique de Paris, 98 bis Boulevard Arago, Paris 75014, France}


\begin{abstract}
By using cosmological hydrodynamical simulations we study the 
effect of supernova (SN) and active galactic nuclei (AGN) feedback on the 
mass transport of gas on to galactic nuclei and the black hole (BH) growth 
down to redshift $z\sim6$. We study the BH growth in relation with the mass 
transport processes associated with gravity and pressure torques, and how 
they are modified by feedback. Cosmological gas funelled through cold flows reaches the galactic outer region close to free-fall. Then torques associated to pressure triggered 
by gas turbulent motions produced in the circum-galactic medium by shocks and explosions from 
SNe are the main source of mass transport beyond the central $\sim$ 100 pc. Due 
to high concentrations of mass in the central galactic region, gravitational 
torques tend to be more important at high redshift. The combined 
effect of almost free-falling material and both gravity and pressure torques
produces a mass accretion rate of order ${\rm \sim 1}$ M$_\odot$/yr at 
$\sim$ pc scales. In the absence of SN feedback, AGN feedback alone does 
not affect significantly either star formation or BH growth until the BH
reaches a sufficiently high mass of $\sim 10^6$  M$_\odot$ to self-regulate. 
SN feedback alone, instead, decreases both stellar and BH growth. 
Finally, SN and AGN feedback in tandem efficiently 
quench the BH growth, while star formation remains at the levels set by SN
feedback alone due to the small final BH mass, $\sim$ few $10^5$ M$_\odot$. 
SNe create a more rarefied and hot environment where energy injection from 
the central AGN can accelerate the gas further. 
\end{abstract}


\section{Introduction}

There is dynamical evidence for the existence of super massive BHs in 
the center of many nearby galaxies \citep{Ferrarese&Ford2005} with 
masses in the range $M_{\rm BH}\sim 10^6-10^9$ M$_\odot$ suggesting 
that BHs, formed at the first evolutionary stages of our Universe, 
now are living in the galactic centers around us including our galaxy 
\citep{Ghez+2005}. 

Scaling relations connect the mass of BHs in local galaxies with their 
host galaxy properties, such as, e.g., the galactic bulge mass
\citep[e.g. ][]{Magorrian+1998,Gultekin+2009}, and the bulge stars 
velocity dispersion \citep[e.g. ][]{FerrareseMerritt2000,Tremaine+2002}. 
Such relations suggest a co-evolution between the BH and its host galaxy 
\citep[see for instance][]{Dressler1989,Kormendy&Richstone1995,Magorrian+1998,Gebhardt+2000},
with a recent work suggesting that such a co-evolution can be trigger 
in galactic bulges with a critical mass above $\sim 10^6$ M$_\odot$
in the early stages of the galactic evolution \citep{Park+2016}.

Detection of very bright quasars at redshift $z\ga6$ with luminosities 
above $10^{13}$ L$_\odot$ implies the existence of BH with masses of the order 
$M_{\rm BH}\sim 10^{9}$ M$_\odot$ when our Universe was about $\sim 1$ Gyr old 
\citep{Fan+2001,Willott+2007,Mortlock+2011,DeRosa+2014}, i.e. BHs should have 
formed very early in the history of our Universe and grow rapidly in order to 
attain such high masses after $\sim$ Gyr~\citep{Dubois+2012,DiMatteo+2012,DiMatteo+2016}. 
To understand such a rapid early evolution is one of the main challenges of the 
current galaxy formation theories \citep{Volonteri2010,Haiman2013}.

In previous work, \citet{Prieto&Escala2016} (hereafter PE16) have studied
the mass transport (MT) process in high redshift galactic discs focusing 
on the effect of SN feedback without taking into account AGN activity on 
such objects. They show that SN feedback is able to affect dramatically 
the BH growth in the first galaxies at high redshift 
\citep[see also][]{Dubois+2015,2016arXiv160509394H, Bower+2016}.

As in PE16, we study the evolution of a high redshift galaxy. In particular,
we compute internal dynamical properties of the system to study the effect 
of AGN activity in the MT process in galactic discs at redshift $z\ga 6$. 
Outflows generated by AGN activity can have spatial extensions of at least
$\sim$ kpc and can reach velocities of the order of $\sim 10^3$ km/s
\citep[e.g. ][]{Humphrey+2010,Nesvadba+2011,Arribas+2014,Genzel+2014}. Such 
powerful outflows should be capable of affecting the host galaxy evolution 
as well as the BH growth. We explore the individual effect of SN and AGN 
in the galaxy evolution and, furthermore, we study how super-Eddington 
accretion may affect the system.

The paper is organized as follows. Section \S 2 contains the numerical 
details of our simulations. In section \S 3 we present our results
based on the gas dynamic analysis of our simulations and its effect the
on the BH growth. In section \S 4 we discuss them, and present our 
main conclusions. 

\section{Methodology and Numerical Simulation Details}
\label{Methodology}

The simulations analysed in this paper are an extension of the work 
presented in PE16 with a few modifications. Therefore, we briefly 
recall the physical ingredients of the numerical experiments.

The simulations were performed with the cosmological N-body
hydrodynamical code RAMSES \citep{Teyssier2002}. The code uses adaptive
mesh refinement, and solves the Euler equations with a second-order 
Godunov method and MUSCL scheme using a MinMod total variation diminishing 
scheme to reconstruct the cell centered values at cell interfaces.

Cosmological initial conditions were generated with the MPGRAFIC code 
\citep{Prunetetal2008} inside a $L=10$ cMpc side box. Cosmological 
parameters were taken from \citet{Planck2013Results} with total matter 
density $\Omega_{\rm m}=0.3175$, dark energy density $\Omega_{\rm \Lambda}=0.6825$, 
baryon matter density $\Omega_{\rm b}=0.04899$, reduced Hubble 
parameter $h=0.6711$, amplitude of the power spectrum at scale of 
$8 h^{-1}\,\rm Mpc$ $\sigma_8 =0.83$ and spectral index $n_{\rm s}=0.9624$.

We have selected a $3\times 10^{10}$ M$_\odot$ DM halo at $z=6$ to be 
re-simulated at high resolution from $z_{\rm ini}=100$ to $z_{\rm end}=6$. 
The high-resolution particles have mass $m_{\rm part}\approx3\times 10^4$ 
M$_\odot$ (which corresponds to an effective resolution of $1024^3$ DM particles 
inside the box). Such a mass allows us to resolve our final halo with $\sim 10^6$ 
DM particles. In order to resolve all the interesting regions, we allowed 
refinements inside the Lagrangian patch associated with a sphere of radius 
$R_{\rm ref}=3R_{\rm vir}$ around the selected DM halo at $z_{\rm end}$
(here $R_{\rm vir}$ is the DM halo virial radius, defined as the radius 
associated to a spherical overdensity 200 times that of the mean matter
density of the Universe at the corresponding redshift). 
The Lagrangian volume (the mask) is defined by an additional scalar 
advected passively with the flow throughout the simulation. At the beginning 
the passive scalar has a value equal to 1 inside the mask and 0 outside. 
In regions where this passive scalar is larger than $10^{-3}$ we allow 
refinement in a cell if any of these conditions is met: i) its total 
mass is more than 8 times that of the initial mass resolution, ii) the Jeans 
length is resolved by less than 4 cells \citep{Trueloveetal1997}, iii) if the 
relative pressure variation between cells is larger than a factor of 2. 
Following these criteria, we reach a maximum co-moving spatial resolution of 
$\Delta x_{\rm min}\approx38.1$ cpc and a proper spatial resolution of 
$\Delta x_{\rm min}\approx5.4$ pc at redshift $z_{\rm end}$, whereas the 
coarse grid inside the mask has a resolution of $\Delta x_{\rm coarse}\approx 9.8$ 
ckpc. The gravitational force resolution is $\Delta x_{\rm min}$ throughout 
the simulation.

Our simulations include optically thin (no self-shielding) gas cooling 
following the \citet{SutherlandDopita93} model down to temperature 
$T=10^4$ K with a contribution from  metals assuming a primordial 
composition of the various heavy elements. Below this temperature, 
the gas can cool down to $T=10$ K due to metal lines cooling 
\citep{DalgarnoMcCray72}. We started the simulations with an 
initial metallicity of $Z_{\rm ini}=0.001$ Z$_\odot$ \citep{Powelletal2011}. 
A uniform UV background is activated at $z_{\rm reion}=8.5$, following 
\citet{HaardtMadau1996}. 

We adopted a star formation number density threshold of 
$n_0\approx 30$ H cm$^{-3}$ with a star formation efficiency 
${\rm \epsilon_\star=0.05}$ \citep[e.g. ][]{RaseraTeyssier2006,DuboisTeyssier2008}.
When a cell reaches the conditions for star formation, star particles 
can be spawned following a Poisson distribution  with a mass resolution 
of $m_{\star,\rm res}\approx5\times10^4$ $\rm M_\odot$. In order to 
ensure numerical stability we do not allow to convert more than $50\%$ 
of the gas into stars inside a cell in one time step.

After 10 Myr the most massive stars explode as SN  releasing a specific 
energy of $E_{\rm SN}=10^{50}$ erg/M$_\odot$, returning 10 per cent of 
the stellar particle mass back into the gas and with a yield of $0.1$ 
inside a sphere of $r_{\rm SN}=2\Delta x_{\rm min}$. As in PE16 we used 
the delayed cooling implementation of SN feedback \citep{Teyssier+2013}. 
It means that, where SNe explode, if the gas ``non-thermal''
energy (stored in a separate passive variable) is above an energy threshold $e_{\rm NT}$, gas cooling is turned
off in order to take into account the unresolved 
chaotic turbulent energy source of the explosions. 
This non-thermal energy component dissipates energy at its own rate with characteristic time $t_{\rm diss}$.
In this work (as in PE16)
we use $t_{\rm diss}=0.5$ Myr and the energy threshold $e_{\rm NT}$ is the one
associated to a turbulent velocity dispersion $\sigma_{\rm NT}=50$ km/s
through the equation $e_{\rm NT}=\rho\sigma_{\rm NT}^2/2$, with $\rho$ the
gas density. Such a velocity dispersion is appropriate for our spatial 
resolution \citep[see][for details]{Dubois+2015,Prieto&Escala2016}.

In order to follow the evolution of the central BH in the simulations, 
we introduced a sink particle \citep{Bleuler&Teyssier2014} when the main 
DM halo has a mass $M_{\rm h}\approx 1.7\times10^{8}$ M$_\odot$ at redshift 
$z=15.7$. The BH seed mass is $10^4$ M$_\odot$. Such a BH mass is in the 
range of masses associated with direct collapse scenario 
\citep[e.g. ][]{OhHaiman2002,LodatoNatarajan2006,Begelman+2006,Begelman+2008,Agarwal+2012,Latif+2013,Latif+2014,Choi+2015}. 
Only this one single BH is allowed to form in the simulation. 

In order to compute the mass accretion rate onto the BH we use the standard
Bondi-Hoyle \citep{Bondi1952} implementation of the accretion rate, $\dot{M}_{\rm Bondi}$. 
In all our simulations expect for one, we cap the accretion rate at 
the Eddington luminosity. For more details on the numerical 
implementation of the processes described above, see PE16.

In order to avoid spurious oscillations in the sink position we have
introduced a drag force acting on the BH particle (Biernacki et al., in 
prep.). The drag force comes from a new source of momentum acting on the 
sink. Such a momentum variation is proportional the cell gas-sink particle
relative velocity. Furthermore, it depends on the non accreted mass
above the Eddington limit: when the BH accretion rate is below the Eddington
limit the drag force is null, it works only if the Bondi BH mass accretion rate 
is above the Eddington limit. In this sense it can be interpreted as a 
force associated to the Eddington pressure around the BH particle. 
In the simulation 
where accretion is not capped at the Eddington limit there is no
drag force. However, even in this extreme case the sink particle stays 
at the center of the galaxy.

We have also  included AGN feedback from the BH.  AGN feedback is modeled 
with thermal energy input \citep{Teyssier+2011,Dubois+2012agn}. The rate of 
energy deposited by the BH inside the injection radius 
$r_{\rm inj}\equiv 4\Delta x_{\rm min}$ is
\begin{equation}
\dot{E}_{\rm AGN}=\epsilon_{\rm c}\epsilon_{\rm r}\dot{M}_{\rm BH}c^2.
\end{equation}
In the above expression, $\epsilon_{\rm r}=0.1$ is the radiative efficiency 
for a standard thin accretion disc \citep{SS1973} and $\epsilon_{\rm c}=0.15$ 
is the fraction of this energy coupled to the gas in order to reproduce the 
local BH-galaxy mass relation~\citep{Dubois+2012agn}. As explained in 
\citet{BoothSchaye2009}, in order to avoid  gas over-cooling  the AGN 
energy is not released instantaneously every time step $\Delta t$ but 
it is accumulated until the surrounding gas temperature can be increased by
 $\Delta T_{\rm min}=10^7$ K. The time between the energy injection events 
depends on the BH mass, mean gas density, and the mass accretion 
rate. It is $\Delta t_{\rm AGN}\sim 1-100$ kyr in our simulations.

\section{Results}
\label{results}

In this work we will show results from four simulations: 
\begin{itemize}
\item SNe run: it includes star formation, SN feedback, modified 
Bondi-Hoyle-Lyttleton (BHL) accretion rate onto sinks without AGN 
feedback. This case was extensively analyzed
in PE16 and here is used to compare it with our AGN simulations. 
\item SNeAGN run: same as SNe plus AGN feedback.
\item NoSNeAGN run: same as SNeAGN but without SN feedback, and
\item SNeAGNnoEdd run: same as SNeAGN but without capping at the Eddington 
limit. 
\end{itemize}

\begin{figure}
\centering
\includegraphics[width=18cm,height=18cm]{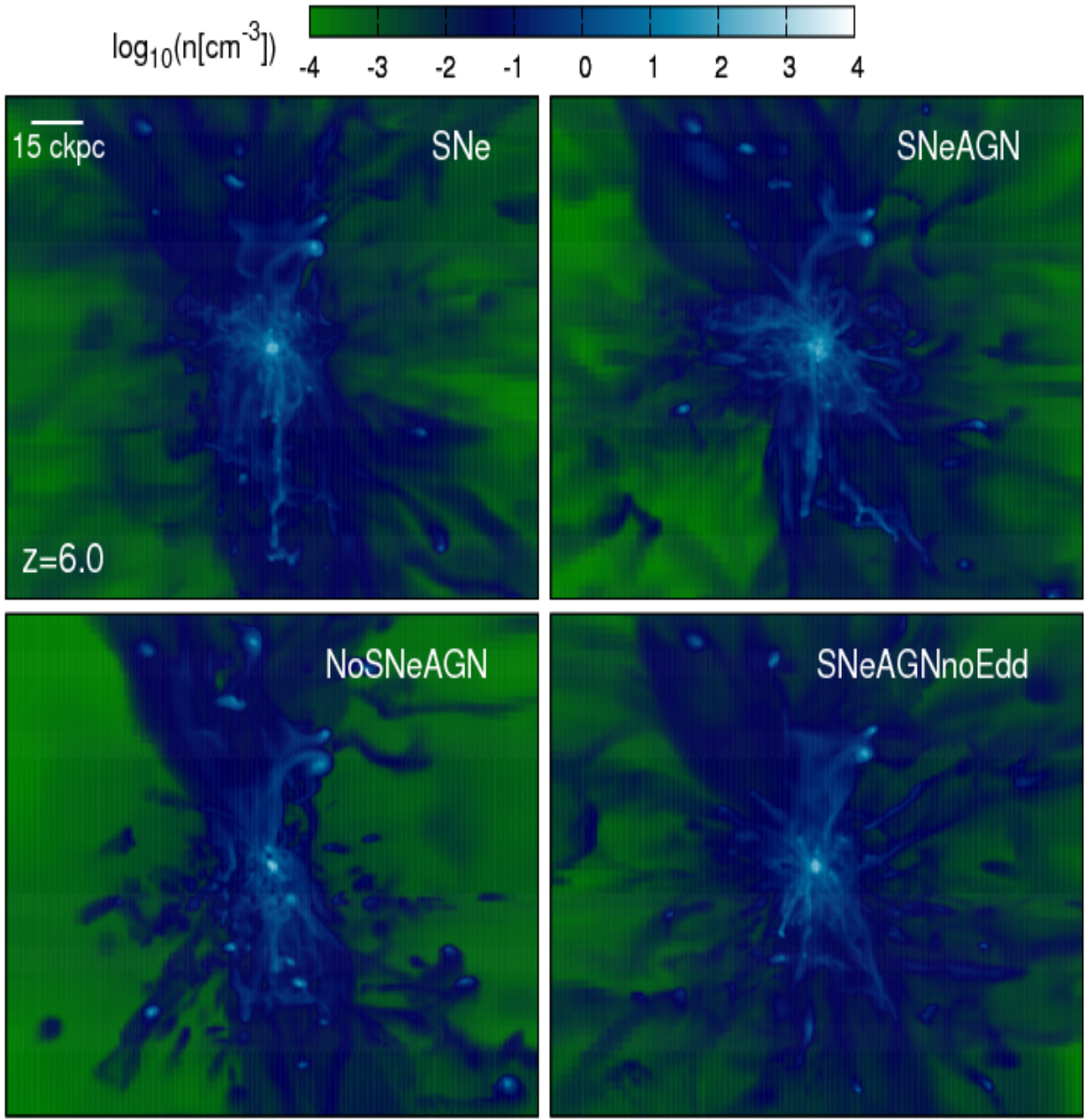}
\caption{Gas number density projections for the four experiments at the end
of the runs. Each panel has an extension of $\sim$ 120 ckpc. It is interesting 
to note that the SNeAGN experiment presents a galaxy much more extended 
compared with the NoSNeAGN simulation where a dense central knot is surrounded 
by low density gas. Such a morphological difference shows how important is the 
SNe heating to create a rarefied environment where the AGN feedback can work 
much more efficiently.}
\label{f1}
\end{figure}

Figure \ref{f1} shows the gas number density projection for the four simulations
at the end of the experiments. Each panel present different features depending 
on the different feedback recipes as will be explained in the following sections.

\subsection{Mass transport on large scales}
Because in a cosmological context at high redshift we can not study the 
small-galactic scale phenomena without taking into account the effects 
of the large scale structure, here we study the behavior of mass 
accretion from $\sim$ few hundreds pc up to $\sim$ 
$3R_{\rm vir}$, with $R_{\rm vir}$ the DM halo virial radius.

\begin{figure}
\centering
\includegraphics[width=18cm,height=18cm]{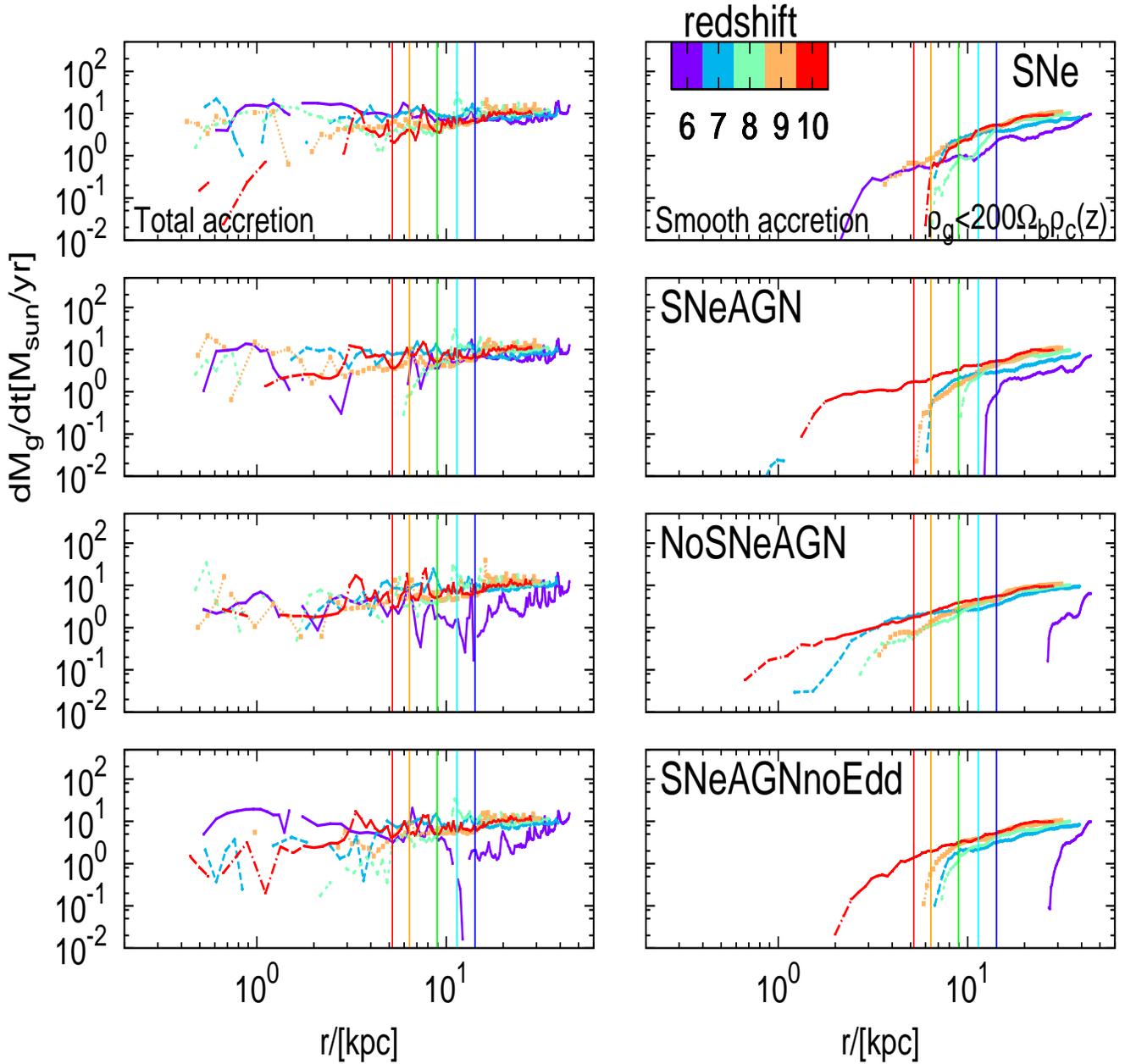}
\caption{Left column: total gas mass accretion rate as a function of 
radius for our different runs at different redshifts, z=10 in red, 
z=9 in orange, z=8 in green z=7 in light-blue and z=6 in blue. From 
top to bottom: SNe, SNeAGN, NoSNeAGN and SNeAGNnoEdd. Right column: 
same as the left column, 
but for low-density smooth accretion, i.e. for gas with a density below 
the collapse density $\rho_{\rm g}< 200\Omega_{\rm b}\rho_{\rm c}$. 
Beyond the virial radius (marked as vertical lines) the total mass 
accretion rate has a floor similar to the smooth accretion: $\sim$ 
few tens of M$_\odot$/yr. Inside the virial radius the accretion 
rate is dominated by dense $n\la 10^3$ cm$^{-3}$ gas.}
\label{f2}
\end{figure}

Fig. \ref{f2} shows the mass accretion rate, computed taking 
into account all the gas mass crossing 
a spherical shell at a given radius centred at the sink cell position:
\begin{equation}
\frac{{\rm d}M_{\rm g}}{{\rm d}t}=-4\pi r^2\rho v_r\, ,
\label{accrate}
\end{equation}
where $\rho$ is the gas mass density and $v_r$ is the total radial gas 
velocity.

The left column of Fig. \ref{f2} shows the total gas mass 
accretion rate for our four simulations as a function of radius. The 
right column shows instead the mass accretion rate associated with 
low gas density 
$\rho_{\rm coll}=18\pi^2\Omega_{\rm b}\rho_{\rm c}\approx 200\Omega_{\rm b}\rho_{\rm c}$, 
with $\rho_{\rm c}$ the critical density of the Universe. For the redshift range 
shown in the figure, $\rho_{\rm coll}\approx (2-7)\times10^{-2}$ cm$^{-3}$. 
This range of densities is below the mean baryon density 
in cold flows in our simulations, which is $\sim 1-0.1$ cm$^{-3}$. The 
vertical lines mark the DM virial radius at each sampled redshift.

Beyond the virial radius all our simulations show a mass accretion 
rate ${\rm \sim 10}$ M$_\odot\,\rm yr^{-1}$ in agreement with PE16. 
This mass accretion rate is associated to gas almost free falling on 
to the DM halo. At these radii the inflowing material is dominated by 
gas with densities $n\sim 1-0.1$ cm$^{-3}$. Those are characteristic 
densities in cooling flows. These quantity shows peaks
($\sim$ few $10^1$ M$_\odot\,\rm yr^{-1}$) associated to both gas clumps 
and DM mini haloes inside the virial radius. From the right column, 
we see how feedback suppresses low density ($n\sim 10^{-2}$ cm$^{-3}$) 
gas accretion inside the inner $\sim$ kpc. Because of the feedback 
heating, only the densest gas is able to flow into the galactic 
central region. At the end of the simulations, gas with densities 
$n\sim 1-0.1$ cm$^{-3}$ can reach the outer galactic region 
($\sim 0.1R_{\rm vir}$) but not beyond, because it is heated and expelled 
from the galaxy. When we include material with density below 10 cm$^{-3}$, 
it can penetrate the galaxy, but it cannot reach the central $\sim$ few 
$100$ pc. Only the densest gas can reach the galactic central region: 
material with density below $\sim$ 100 cm$^{-3}$ is able to reach the 
central $\sim$ few 100 pc region. 

The AGN simulations show that at $z=6$ the low density material can 
not penetrate inside the virial radius. By this time, AGN activity is 
capable to evaporate the diffuse material. At higher redshift the diffuse 
gas can reach smaller radii but the accreted material is dominated by 
high density gas. In particular, our NoSNeAGN simulation has the deepest 
low density gas penetration at $z\geq 7$. In this case due to the no SN 
heating and the low BH mass (few $10^6$ M$_\odot$), AGN feedback is unable 
to alter significantly the low density $\sim$ kpc scale accretion. At lower 
redshift the BH experiences a prolonged Eddington limited accretion rate 
(as discussed in the following) that increases its mass by a factor of $\sim$ 
a few, injecting in turn energy in the interstellar and circum-galactic media.
Such energy injection clearly affects the low density gas accretion rate at 
$z=6$, as we can see from the figure: the low density gas barely reaches the
virial radius.

We note that the expression in eq. \ref{accrate}
includes mass inflows associated to inward radial motions triggered 
by local thermal fluctuations. In order to asses how important are the
local thermal fluctuations to the mass accretion rate,
we compute the radial velocity dispersion $\sigma_{r,{\rm NoSNe}}$ in 
the no feedback simulation of PE16. Then, for each simulation presented 
in this work, we compute the mass accretion rate including all the 
material with radial velocity 
\begin{equation}
|v_r| > \sqrt{\sigma_{r,{\rm NoSNe}}^2},
\end{equation} 
i.e. we excluded local thermal fluctuations. Following this procedure we can 
see that at $\sim$ virial radius scales the local velocity fluctuations 
are no more than $\sim 20\%$ of the total accreted matter.

The total mass accretion rate in our four simulations has similar values at 
$\sim$ kpc scales fluctuating between $\la 10^{-1}$ M$_\odot\,\rm yr^{-1}$ and 
$\sim$ few tens of M$_\odot\,\rm yr^{-1}$, showing that the effect of the AGN
activity has not a notable impact on $\sim$ kpc scales in these small high
redshift galaxies, due to the small BH mass. Such accretion rates of order 
$\sim$ few tens of M$_\odot\,\rm yr^{-1}$ at the outer galactic edge are
associated to free fall material reaching the central DM halo region from 
large scales, as discussed in PE16.

\subsection{Mass transport in the disc}
The tight relation between the large scale $\ga R_{\rm vir}$ dynamic and the
small scale $\la R_{\rm vir}$ evolution in the galaxy formation context
\citep[e.g. ][]{Pichon+2011,Dubois+2012,PrietoSpin,Danovich2015} is a 
robust motivation to study the MT process at different scales in the first
galaxies. In the following, we analyze the MT process on $\la$ kpc 
($\la 0.1R_{\rm vir}$) scales. 

\subsubsection{Torques in the disc}
After to flow at almost free fall from scales above $\sim R_{\rm vir}$ till 
the galactic edge (at $\sim 0.1R_{\rm vir}$) triggered by gravity and
channelized through DM filaments around the DM halo, the angular momentum 
(AM) re-distribution in the galaxy will produce MT, allowing the gas to flow 
in and to reach the center of the system. Due to the nature of the system 
at study, the sources of AM variations are related to gravitational forces and 
pressure gradients, namely:
\begin{equation}
\vec{\tau}_{\rm G}=\vec{r}\times \nabla\phi,
\end{equation}
\begin{equation}
\vec{\tau}_{\rm P}=\vec{r}\times\frac{\nabla P}{\rho}.
\end{equation}
The gravitational, $\vec{\tau}_{\rm G}$, and the pressure gradient, 
$\vec{\tau}_{\rm P}$, torques will act as source of AM transport in the galactic 
system and provide clues about the MT process in high redshift galaxies. 

\begin{figure}
\centering
\includegraphics[width=18cm,height=18cm]{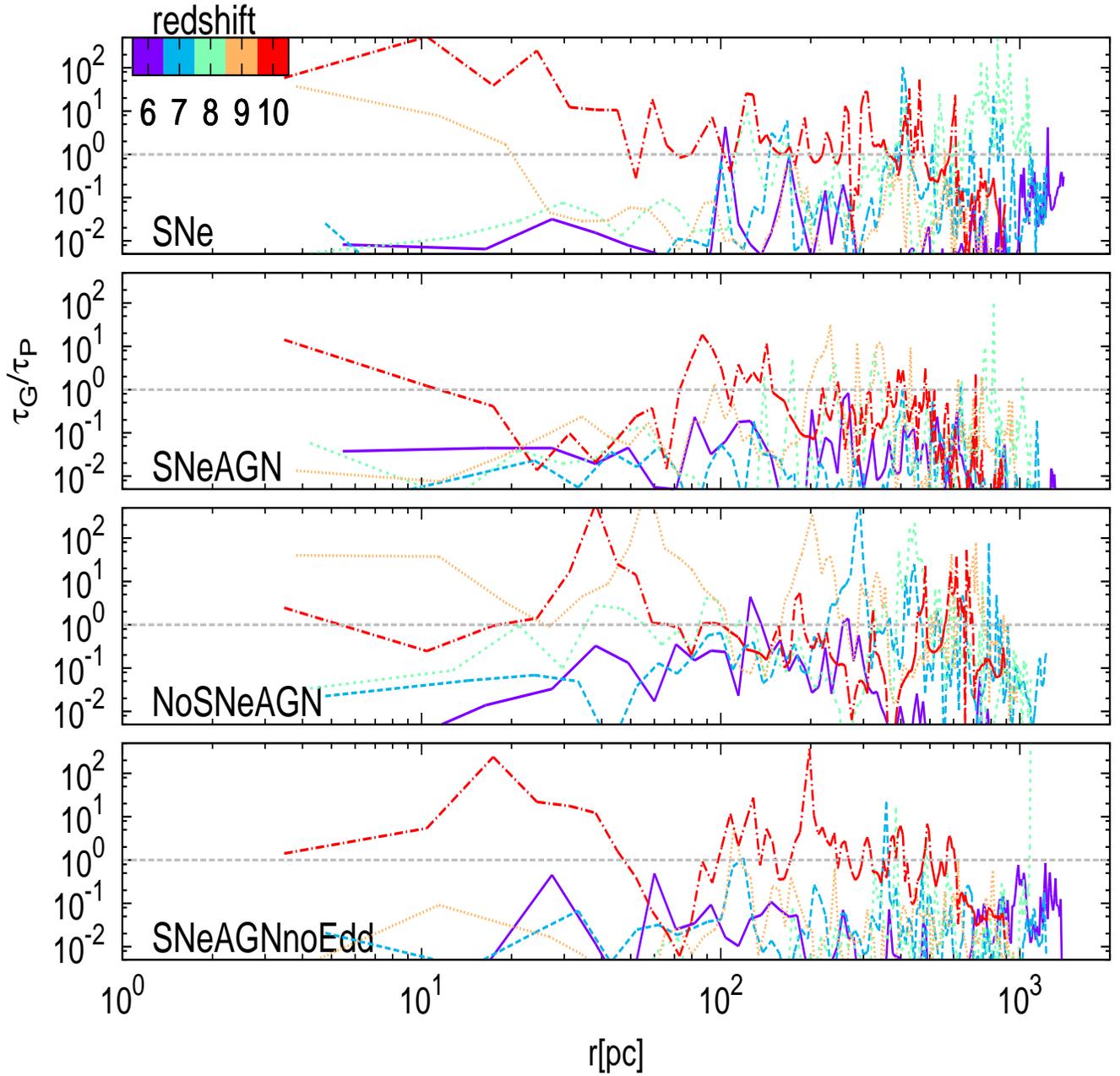}
\caption{Gravitational torque to pressure gradient torque ratio 
as a function of radius for different redshifts: z=10 in red, 
z=9 in orange, z=8 in green z=7 in light-blue and z=6 in blue.
The ratio is smoothed over $\sim 30$ pc in radius. 
From top to bottom: SNe, SNeAGN, NoSNeAGN and SNeAGNnoEdd. 
The gray dashed line marks the $\tau_{\rm G}/\tau_{\rm P}=1$ 
state. Above $\sim$ 100 pc the systems tend to be dominated by 
pressure gradient torques. Despite of that  gravitational torques 
are also effective at ${\rm r\ga 100}$ pc at some regions.
The no SN feedback simulations shows the higher gravitational 
contribution below ${\rm r\ga 100}$ due to the lack of SN heating.}
\label{f3}
\end{figure}

Fig. \ref{f3} shows the ratio $\tau_{\rm G}/\tau_{\rm P}$ as 
a function of radius at different redshifts for our four simulations, with 
$\tau_i\equiv |\vec{\tau}_i|$. The data is smoothed over $\sim 30$ pc in radius.
All our simulations show that pressure 
gradient torques tend to dominate above $\sim$ 100 pc, with a clear domination 
at almost all radii at lower redshift. This is the consequence 
of two processes: shocks due to the large scale in-falling gas onto the DM halo
central region and shocks due to SNe and AGN feedback. There are, however, 
some regions of gravity domination above $\sim$ 100 pc.
The pressure gradient domination is clearer at smaller radii. Only at high
redshift ${\rm z\ga 9}$ gravity dominates the center of the system. 
In the case without SN feedback the gravitational 
torque has a contribution at smaller radii ($r\sim$ few tens pc) compared 
with the SN feedback simulations. This is due to the higher mass 
concentration in this simulation. In summary, the combined effect of 
gravity and pressure gradients re-distributes AM in the disk and triggers 
MT that feeds the central BH in these systems. We note that due to the mass
difference (above $\sim$ one order of magnitude in DM halo) our results 
does not completely agree with the ones presented in \citet{Danovich2015}.
These authors show that the gravitational torque dominates on the galactic
disc. As mentioned above such a difference arises due to the bigger mass 
of their studied systems at lower redshift.

\subsubsection{Mass accretion rate in the disc}
As shown in the previous section, high redshift galaxies receive almost 
free falling material from large scale and experience torques associated 
with gravitational forces and pressure gradients which trigger MT in the 
galaxy. In this context, it is relevant to study and quantify the mass
accretion rate in the galactic disc due to such phenomena and on to the 
central BH. 

\begin{figure}
\centering
\includegraphics[width=18cm,height=18cm]{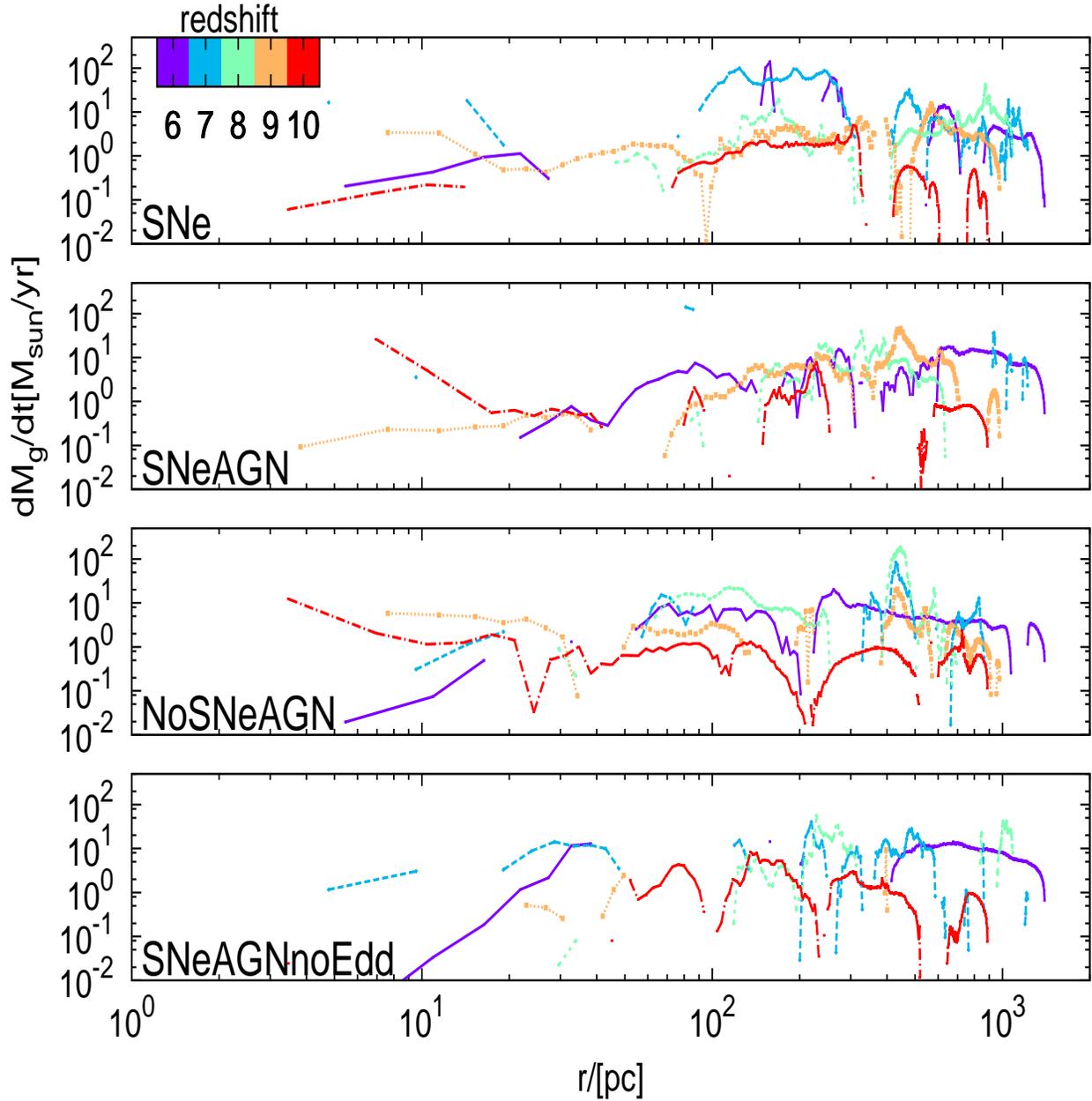}
\caption{Mass accretion rate radial profiles inside $0.1R_{\rm vir}$ 
for our four runs at different redshifts: $z=10$ in red, $z=9$ in 
orange, $z=8$ in green, $z=7$ in light-blue and z=6 in blue. All 
simulations show fluctuations in the mass accretion rate with 
values between $\sim$ $10^{2}$ ${\rm M_\odot/yr}$ and 
$\sim$ $10^{-2}$ ${\rm M_\odot/yr}$. Due to the lack of SN heating 
the NoSNeAGN run has the smoother data at high redshift.}
\label{f4}
\end{figure}

Fig. \ref{f4} shows the radial gas mass accretion rate in the disc 
as a function of radius inside $\sim 0.1R_{\rm vir}$ at different redshifts 
for our four runs. We defined the mass accretion in the disc as:
\begin{equation}
\frac{{\rm d}M_{\rm g}}{{\rm d}t}=-2\pi r\Sigma_{\rm g} v_{\rm r}\, . 
\end{equation}
We first compute the gas AM vector inside a radius $r=0.1 R_{\rm vir}$, 
and take it as the $\hat{z}$ vertical axis of the cylindrical coordinate 
system of reference. The cylindrical radial coordinate $r$ is defined in 
the disc plane perpendicular to the 
AM vector, and both the gas surface density $\Sigma_{\rm g}$ and 
the radial velocity $v_{\rm r}$ are cylindrical shell averages in the 
$\hat{z}$ direction, up to a height such that the $90\%$ of the baryonic mass
is enclosed.

All simulations show large fluctuations in the mass accretion rate, 
attesting to the dynamical conditions in these environments: large-scale 
gas inflows and SN feedback shock the gas creating a turbulent 
environment.  Accretion rates fluctuate between $\sim$ $10^{2}$ 
M$_\odot\,\rm yr^{-1}$ and $\sim$ $10^{-2}$ M$_\odot\,\rm yr^{-1}$ with an
average accretion rate of the order $\sim (1-10)$ M$_\odot\,\rm yr^{-1}$. The 
accretion rate profiles show a number of gaps associated to SN outflows and gas 
clumps crossing and leaving the system.

The lack of SN heating in the NoSNeAGN simulation causes smoother accretion
rate profiles compared with all the SN feedback simulations. Furthermore, 
in simulations with AGN feedback gas does not easily reach the central 
galactic region, $r\la 10$ pc, i.e. AGN feedback efficiently ejects gas 
from the galactic center as we will discuss in the next section.

\subsection{BH evolution}
We have shown that after to flow almost radially at free fall from scales 
above $\sim R_{\rm vir}$ and reach the galactic outer region, 
gravitational and pressure gradient torques can produce a 
substantial mass accretion rate of the order $\sim (1-10)$ M$_\odot$/yr 
at distances $\ga$ few 10 pc from the center of the system in high 
redshift galaxies. We now move to explore the BH mass growth and how 
dynamical features of the system depend on SN and AGN feedback.

\subsubsection{BH accretion rate}
\begin{figure}
\centering
\includegraphics[width=1.0\columnwidth]{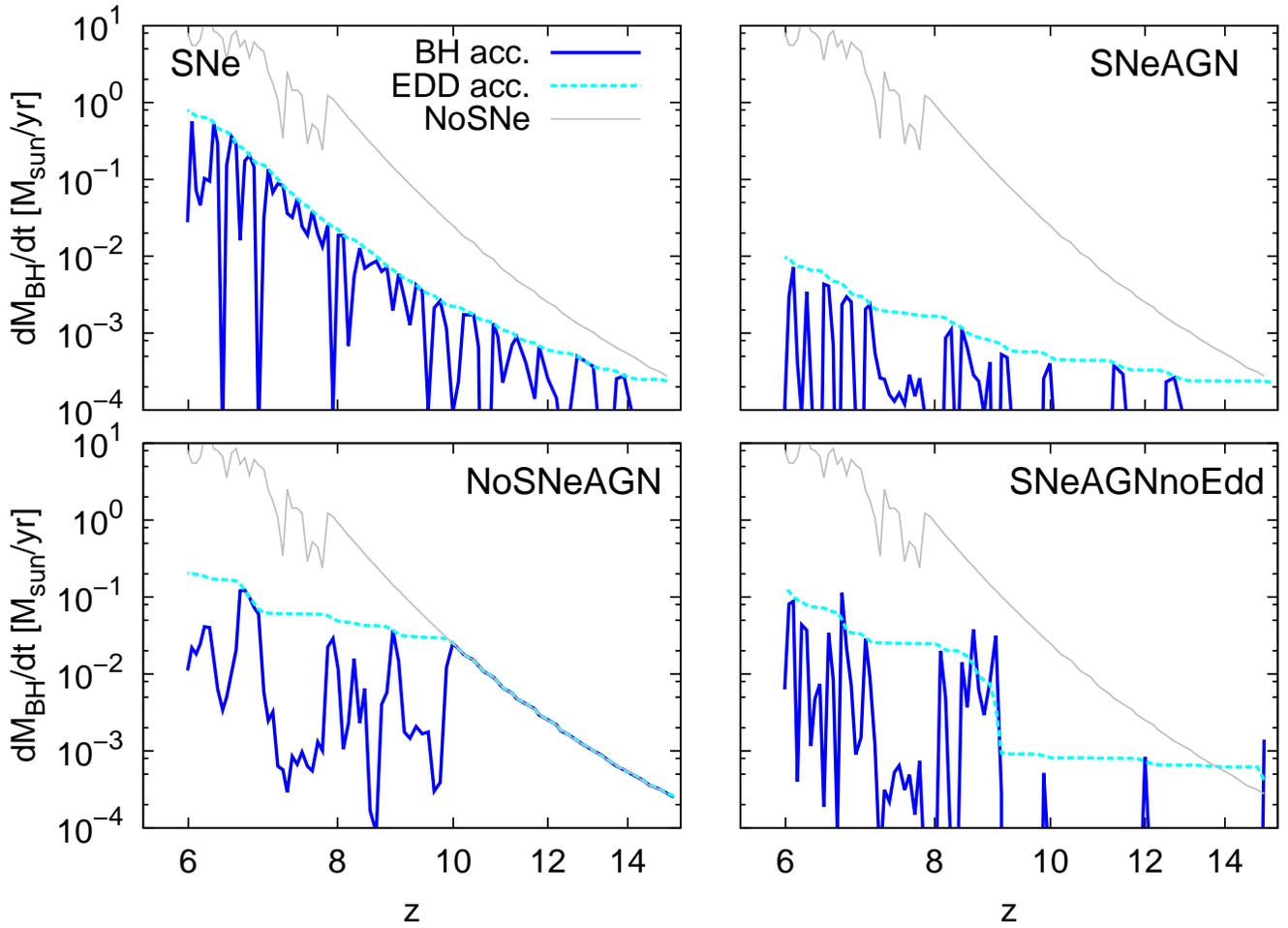}
\caption{BH mass accretion rate for our four simulations as a 
function of redshift is shown with a solid blue line. The dashed 
cyan line marks the Eddington limit. For comparison, the solid gray 
line marks the BH accretion rate of the no feedback NoSNe run in PE16. 
The top left panel clearly shows that SN feedback affect the mass 
accretion rate onto the BH. AGN feedback is able to reduce even 
more the accretion rate compared with the simulation with only 
SN feedback. The NoSNeAGN experiment allows a mass concentration 
in the galactic central region at high redshift producing a 
practically Eddington limited growth of the BH. The SNeAGNnoEdd 
run produces long periods of almost null accretion after 
super-Eddington episodes.}
\label{f5}
\end{figure}

We show the BH accretion rate as a function of redshift in 
Fig.~\ref{f5}. As already shown by PE16, the SN explosions have 
a clear effect on the central BH mass accretion rate ejecting gas out of 
system and then decreasing the amount of material that can feed the 
BH throughout the simulation \citep[see also][]{Dubois+2015}. 
This is especially important at $z> 10$, when the halo mass is  
$\la$ $10^9$ M$_\odot$ and the stellar mass $\la$ few 
$10^7$ M$_\odot$. Under such conditions SN explosions can accelerate a fraction
of the central galactic gas beyond the local escape velocity depleting of 
gas the BH neighborhood. The SN run is characterized by intermittent 
Eddington limited accretion episodes with an average 
$\langle \dot{M}_{\rm BH}/\dot{M}_{\rm Edd} \rangle \equiv \langle f_{\rm Edd}\rangle\approx 0.54$ 
and a final BH mass $M_{\rm BH}\approx 3.6\times 10^7$ $\rm M_\odot$ 
(see table \ref{table}). 

AGN and SN feedback in tandem (SNeAGN) reduce significantly the 
mass accretion rate, a factor $\sim 2$ overall, with 
$\langle f_{\rm Edd}\rangle\approx 0.24$ and a final BH mass 
$M_{\rm BH}\approx 4.3\times 10^5$ $\rm M_\odot$. 
Beside the SN energy injection, the local effect of the AGN activity is able 
to eject the already low amount of gas from the galactic center, reducing
dramatically the mass accretion rate. Notwithstanding, the central BH has 
a number of Eddington limited episodes.

\begin{figure}
\centering
\includegraphics[width=1.0\columnwidth]{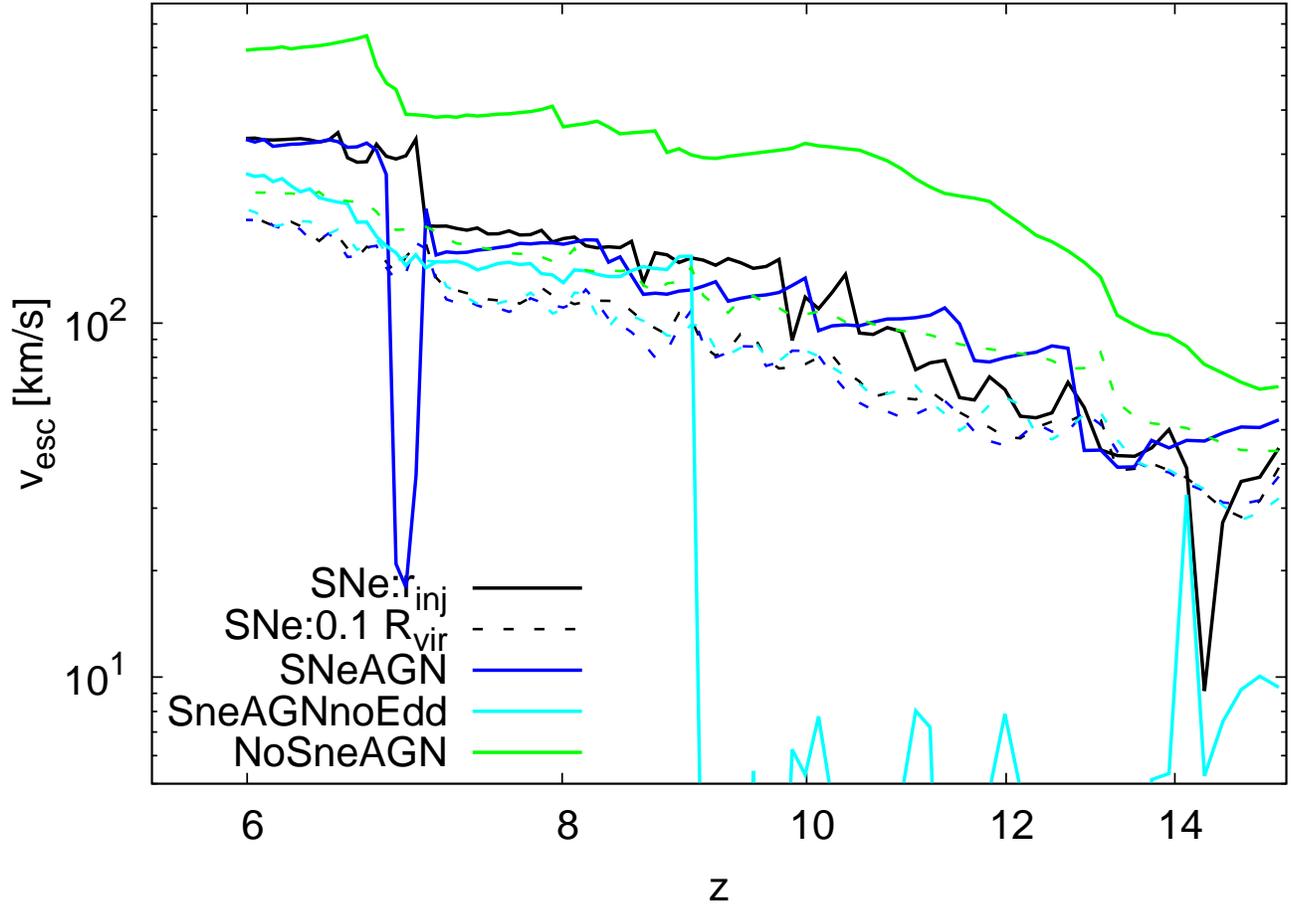}
\caption{Escape speed at different distances from the BH as a 
function of redshift for our four simulations. The solid line 
marks the escape speed at the energy injection radius 
$r_{\rm inj}=4 \Delta x_{\rm min}$ and the dashed line at 
$0.1R_{\rm vir}$. SNe in black, SNeAGN in blue, SNeAGNnoEdd 
in cyan and NoSNeAGN in green. Due to the lack of SN heating 
the no SN feedback experiment has the higher escape speed: it 
allows much more mass concentration in the galaxy. On the other 
hand, the no Eddington limit case depletes of gas the tiny 
galaxy at high redshift after super Eddington episodes, 
decreasing its escape speed.}
\label{f6}
\end{figure}

The NoSNeAGN run shows an interesting behavior. Early on the BH accretes 
close to the Eddington limit as in the no SN feedback case of PE16 (see 
the gray solid line in Fig.~\ref{f5}). This behavior continues to 
$z\sim10$, when the now sufficiently massive BH is capable to accelerate 
the surrounding gas beyond the central escape velocity. The lack of SN 
heating allows the gas to easily reach the galactic center and pile up to 
feed the BH. At the same time the accumulation of gas deepens the 
potential well and increases the central escape velocity, as shown in 
Fig.~\ref{f6}, making it difficult for gas affected by AGN feedback 
to leave the central region. At $z\sim 10$, the BH mass has grown enough 
that AGN feedback becomes sufficiently strong to quench BH accretion (see 
the lower-left panel in Fig.~\ref{f5}). Overall, the mean accretion
rate,  $\langle f_{\rm Edd}\rangle\approx 0.49$, and, final BH mass, 
$M_{\rm BH}\approx 9.3\times 10^6$ $\rm M_\odot$, are in between the cases 
with only SN or both SN and AGN feedback. 

The case without Eddington limit, but with both types of feedback 
included, SNeAGNnoEdd is characterized by an early peak of super-Eddington
accretion at $z\sim 15$. This early burst, in a tiny galaxy, causes
catastrophic feedback that essentially shuts off accretion until $z< 10$. It 
is important to note that feedback has not been modified to take into 
account the lower radiative efficiency expected in super-Eddington accretion
discs \citep[see][and references therein; a simulation following such a scenario is work in progress and will be part of an upcoming paper]{2015ApJ...804..148V}. A lower radiative efficiency decreases the injected energy from
feedback (see Eq.~1) and would be less disruptive on its surroundings 
\citep{2016MNRAS.456.2993L}. In this case the average 
$\langle f_{\rm Edd}\rangle$ is $\approx 0.32$ with a final mass 
$M_{\rm BH}\approx 5.8\times 10^6$ $\rm M_\odot$.

Table \ref{table} shows $\langle f_{\rm Edd}\rangle$ for all our simulations,
for the whole duration of the simulation, as well as only at $z>10$ and 
$z<10$ to highlight the effects described above. In all simulations including
SN feedback, the high redshift ($z>10$) $\langle f_{\rm Edd}\rangle$ is 
lower than the low redshift ($z<10$) $\langle f_{\rm Edd}\rangle$. Between
redshift 6 and 10 the  system has become more massive and SN feedback alone
cannot unbind the gas. In the simulation without SN feedback, but with AGN
feedback, the mean Eddington ratio is higher at higher redshift ($z>10$): AGN
feedback affects the BH accretion rate only when the BH has become sufficiently
massive to drive powerful outflows and clear its immediate surroundings.


\subsubsection{Dynamical conditions}
These simulations show how important outflows are
on the dynamics of the central region of high-redshift galaxies. 
Depending on their power they can change the mass
distribution around the BHs and affect the dynamical 
conditions by changing the local escape velocity. Fig. \ref{f6} 
shows the escape speed for our four simulations at different radii 
as a function 
of redshift. The escape speed from a given radius $R$ is defined by $v_{\rm esc}=\sqrt{2GM(<R)/R}$ with
$M(<R)$ the total mass (BH, gas, stars and DM) inside a radius $R$. The
NoSNeAGN simulation has the largest escape speed 
at the injection radius $r_{\rm inj}=4 \Delta x_{\rm min}$  (solid green
line, this is the velocity needed for the gas to leave the region from which the BH accretes, not for gas to leave the galaxy or the halo) 
and at $0.1 R_{\rm vir}$ (dashed green line, this can be 
considered the velocity needed to leave the galaxy). Due to the lack of SN 
heating, cold dense gas can pile up in the galaxy and in the BH vicinity
increasing the escape velocity. Such conditions clearly favor a high accretion
rate onto the BH until it reaches a sufficiently high mass for its AGN 
activity to impact its surroundings.

In contrast, the SNeAGNnoEdd simulation has the lowest escape speed above 
${\rm z\approx 9}$ at the injection radius. In this case the strong AGN feedback
associated to the early bursts at super-Eddington rates expels the gas 
around the BH, creating a very low density environment with a 
low escape speed (see bottom right panel of Fig. \ref{f5}), which
favors mass depletion in the BH vicinity, and keeps the accretion to minimal
levels. At $z\approx 9.5$ the escape speed increases at similar (but still 
lower) values compared with the other runs. At this redshift the main DM halo
merges with a galaxy which has not been affected by AGN activity, increasing
the enclosed central mass and consequently the escape velocity.

The SNe and SNeAGN runs show similar escape speeds at $0.1 R_{\rm vir}$. 
Their escape speeds have clearer differences at the injection radius. They 
fluctuate due to the SN and BH energy injection which in both cases heat up and 
reduce the amount of available gas in the galactic center. Despite the similar
escape speed in these two simulations, their BH growth 
is rather different (see Fig.~\ref{f5}). 
In these simulations the mass inside $r_{\rm inj}$
is dominated by the stellar component, while accretion depends on gas content
around the BH (as we will discuss later), which shows rapid and large
fluctuations early on for simulations including SN feedback~\citep{Dubois+2015}.
In the SNeAGN simulation, beside the SN mass outflows, AGN feedback is also at 
work, reducing substantially the BH growth rate compared with the BH mass 
accretion rate in the SN only simulation.  

\begin{figure}
\centering
\includegraphics[width=18.5cm,height=18.5cm]{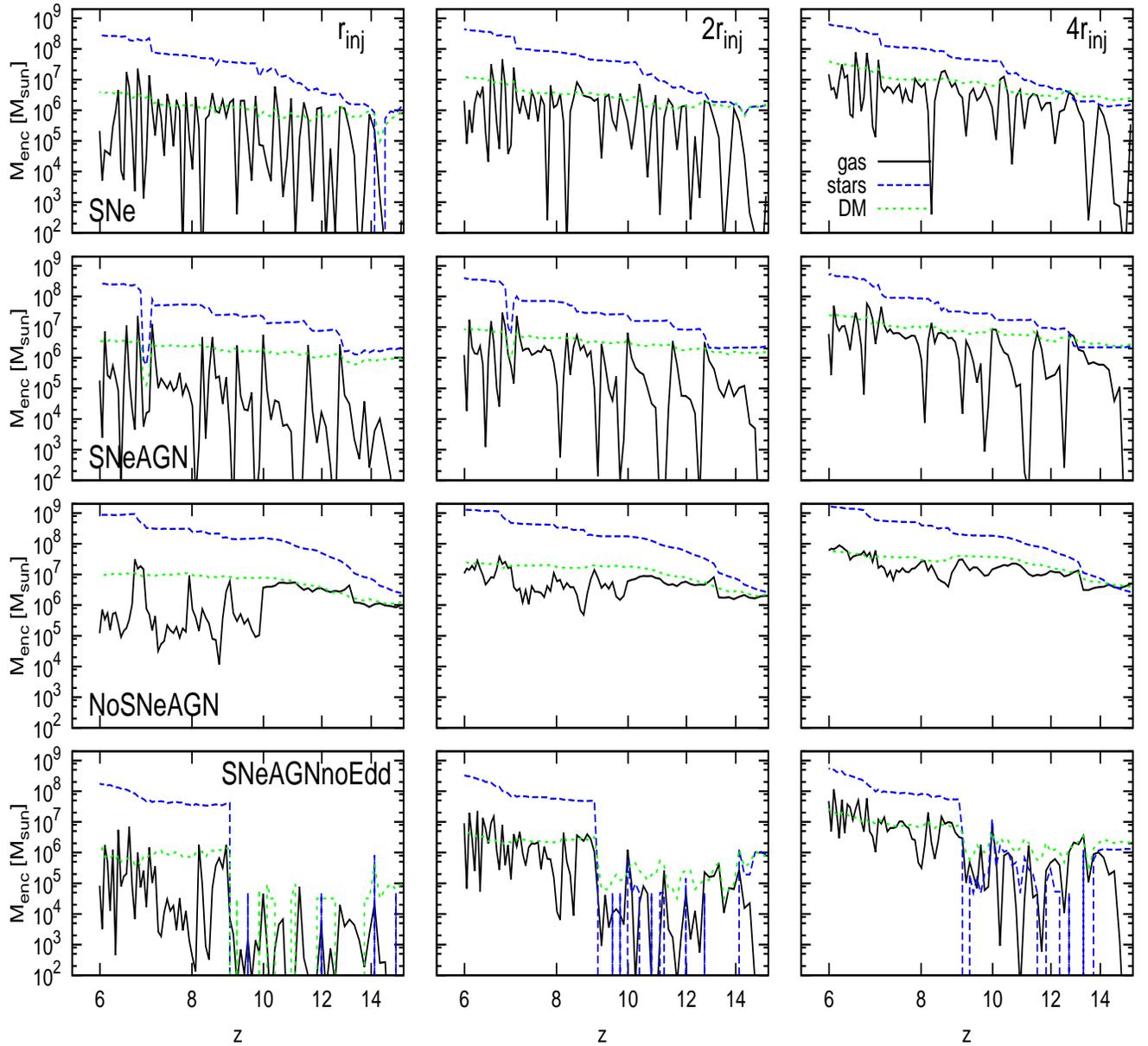}
\caption{Gas (solid black line), stars (long-dashed blue line) and
DM (short-dashed green line) enclosed mass inside three different
distances from the BH position in different columns. Each row shows
the enclosed mass for each simulation as a function of redshift.
The no SN feedback simulation has 
the highest central mass concentration at high redshift. In 
contrast the no Eddington limit run has the lowest one. All the
SN feedback simulations show an irregular behavior in the 
central gas content as a consequence of the regular gas mass 
removal. In all our simulations the central galactic region 
is dominated by the stellar mass except in the no Eddington 
limit case, where the strong early feedback expel the nuclear gas
and suppresses the star formation efficiently.}
\label{f7}
\end{figure}

Figure \ref{f7} shows the gas, stars and DM enclosed mass as a function 
of redshift for all simulations at three different distances from
the BH position, namely $r_{\rm inj}$, $2r_{\rm inj}$ and $4r_{\rm inj}$.
In all these simulations baryons dominate the enclosed mass in the BH
vicinity, with a major contribution from the stellar component as mentioned
above. This figure can explain why the SNeAGN experiment has a lower BH 
accretion rate compared with the only SN feedback run despite their 
escape speed being similar: the gas content around the BH in the AGN feedback
simulation is lower than the one around the BH in the only SNe simulation. 
Only the no-Eddington limited experiment shows that  DM dominates 
the central galactic region at high redshift, $z\ga 9$. In this particular case,
the extreme BH feedback is strong enough to expel the central galactic gas 
and inhibit  star formation.

\begin{figure}
\centering
\includegraphics[width=1.0\columnwidth]{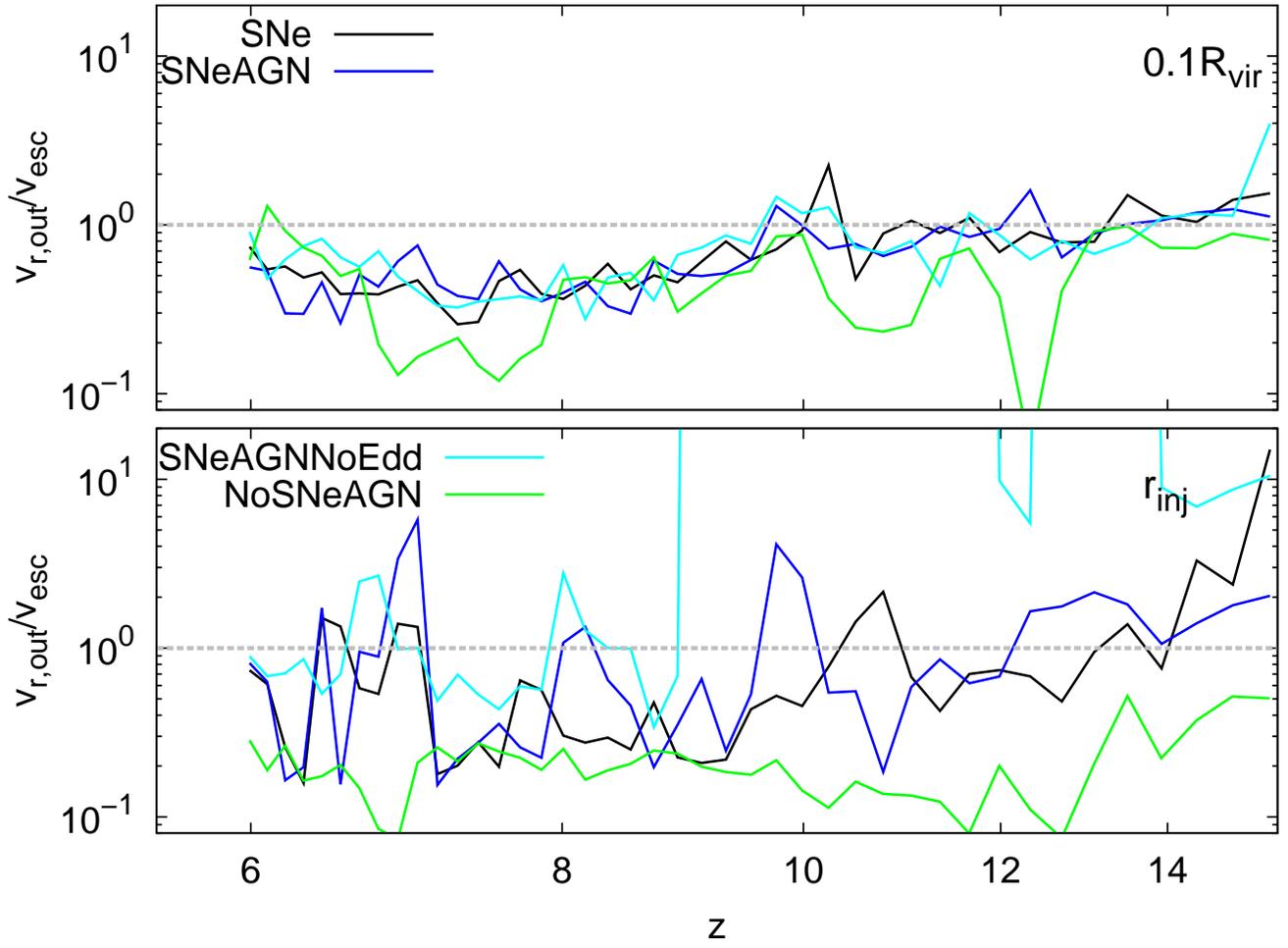}
\caption{Outflow radial velocity normalized to the escape speed 
at different radii: $0.1R_{\rm vir}$ at center and $r_{\rm inj}$ 
at bottom. All ratios are smoothed over $\sim 20-30$ Myr. 
Ejection of gas from the galaxy (with $0.1R_{\rm vir}$ 
as a proxy) is intermittent and not very frequent. Near the BH 
($r_{\rm inj}$), the no SN feedback simulation has the lowest 
velocity ratio, meaning that gas is retained in the center, 
while the simulation without Eddington limiter retains no central gas.}
\label{f8}
\end{figure}

Fig. \ref{f8} shows the average mass weighted gas outflow speed 
normalized by the escape speed, $v_{\rm r,out}/v_{\rm esc}$, as a function of 
redshift for our 
four simulations at two different radii, namely $0.1R_{\rm vir}$ and
$r_{\rm inj}$. The data is smoothed over $\sim 20-30$ Myr.
The dashed gray line shows $v_{\rm r,out}/v_{\rm esc}=1$.
At the injection radius the NoSNeAGN feedback simulation (green line) creates 
outflows with speeds well below the escape speed due to the high mass 
concentration (see Fig.~\ref{f7}) around the BH as mentioned 
above. At larger radii all the SN feedback simulations have an approximately 
similar behavior, showing that the stronger effects can be seen in the 
BH vicinity.  

The run without Eddington limit (cyan line) 
produces nuclear outflows with speeds well above the escape speed at $z\ga 9.5$ 
ejecting most of the gas around the BH and quenching efficiently BH growth and 
central star formation. The combined effect of gas ejection and reduced star 
formation decreases the central mass concentration, reducing the escape
velocity. Under such conditions, feedback can easily unbind the gas, and 
only the merger at $z\sim 9.5$ increases the central mass at levels comparable
with the other SN simulation runs.

The SNe and SNeAGN simulations have a similar behavior in the innermost region
of the galaxy. These simulations have a highly fluctuating behavior with 
periods of bound and unbound central gas which is reflected in the BH mass 
accretion rate fig.~\ref{f5}, however the nature of 
the gas that is accelerated in both cases is different as we will see now.

\subsection{Velocity distribution in the galactic gas}

\begin{figure}
\centering
\includegraphics[width=15.5cm,height=7.2cm]{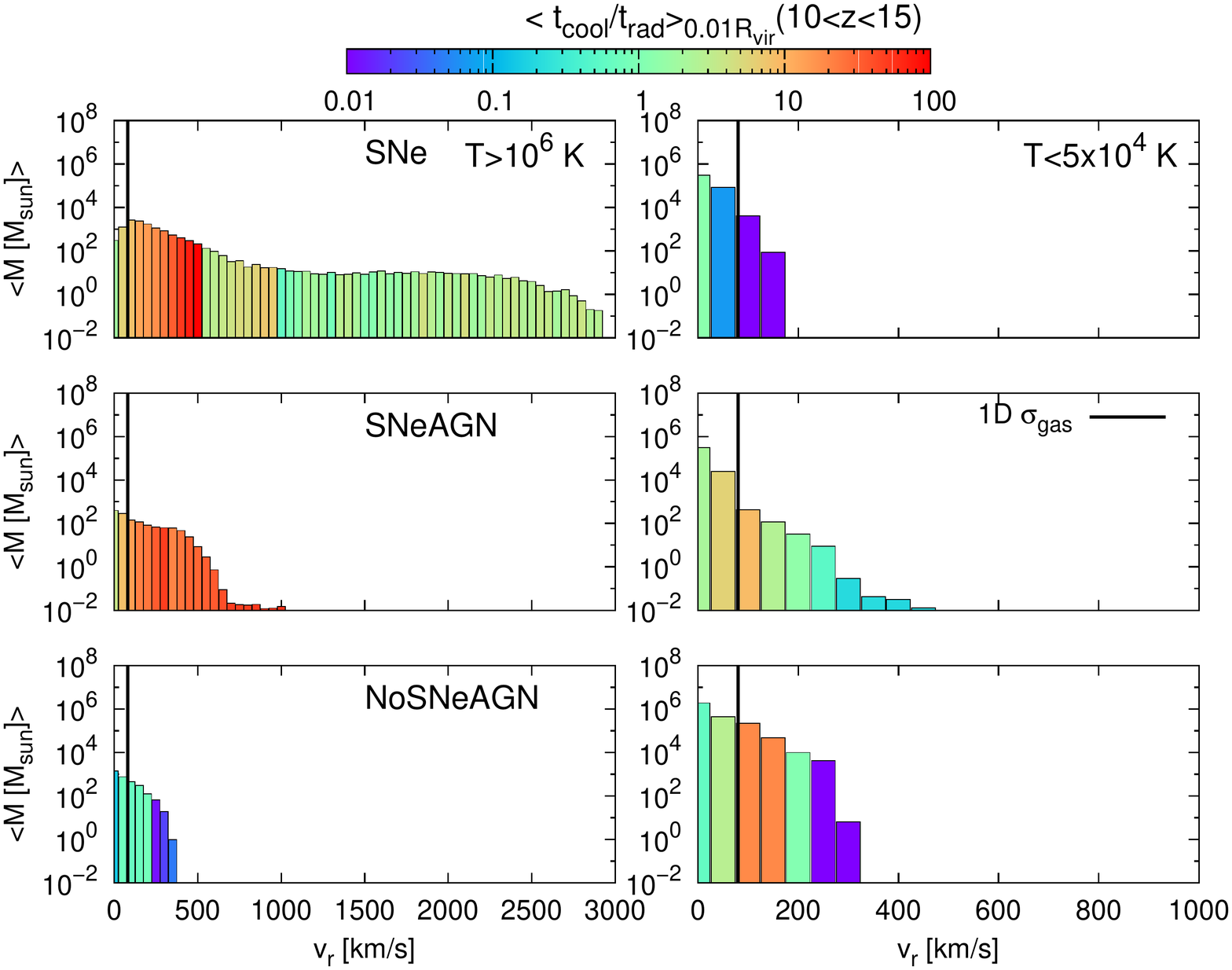}
\includegraphics[width=15.5cm,height=7.2cm]{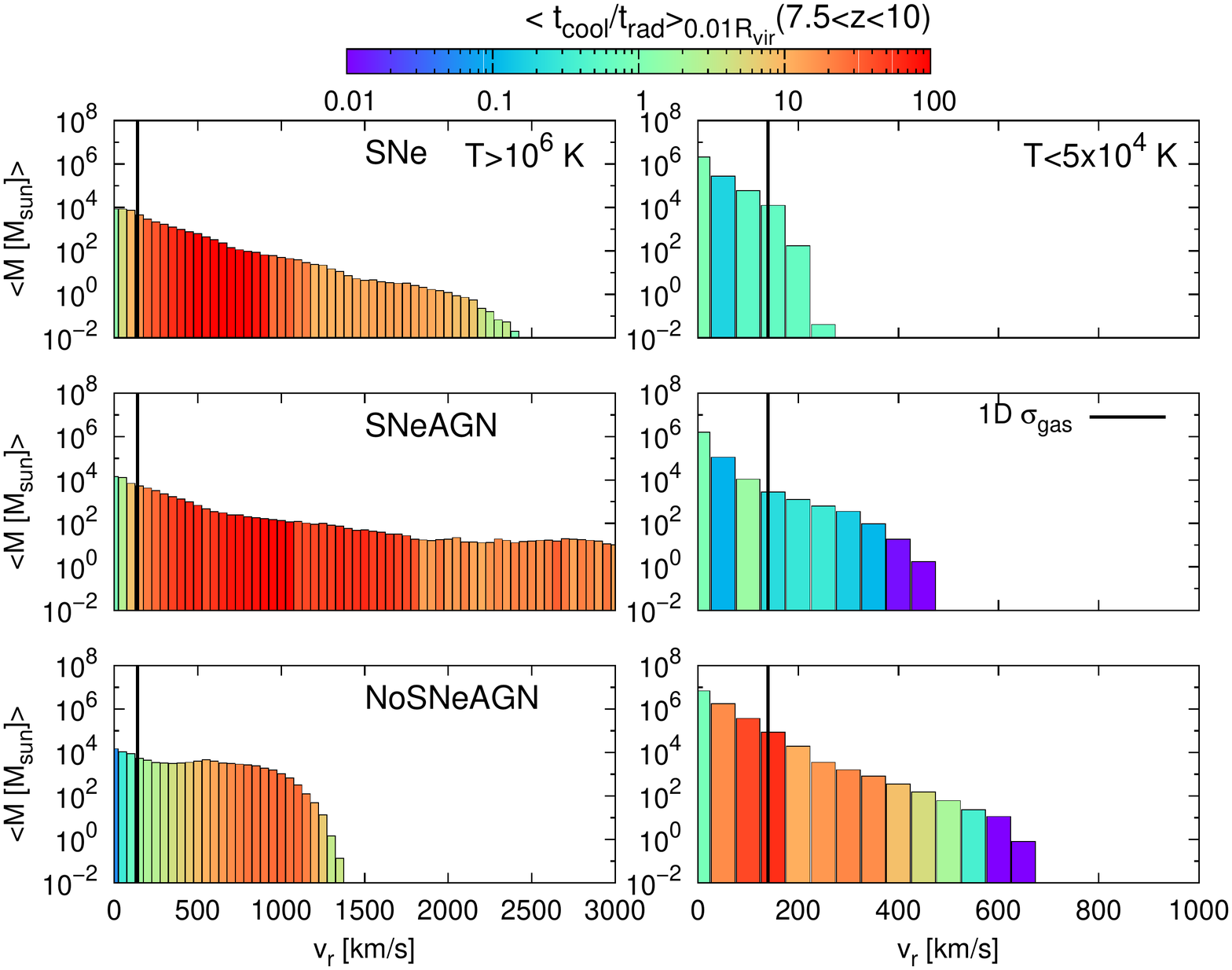}
\includegraphics[width=15.5cm,height=7.2cm]{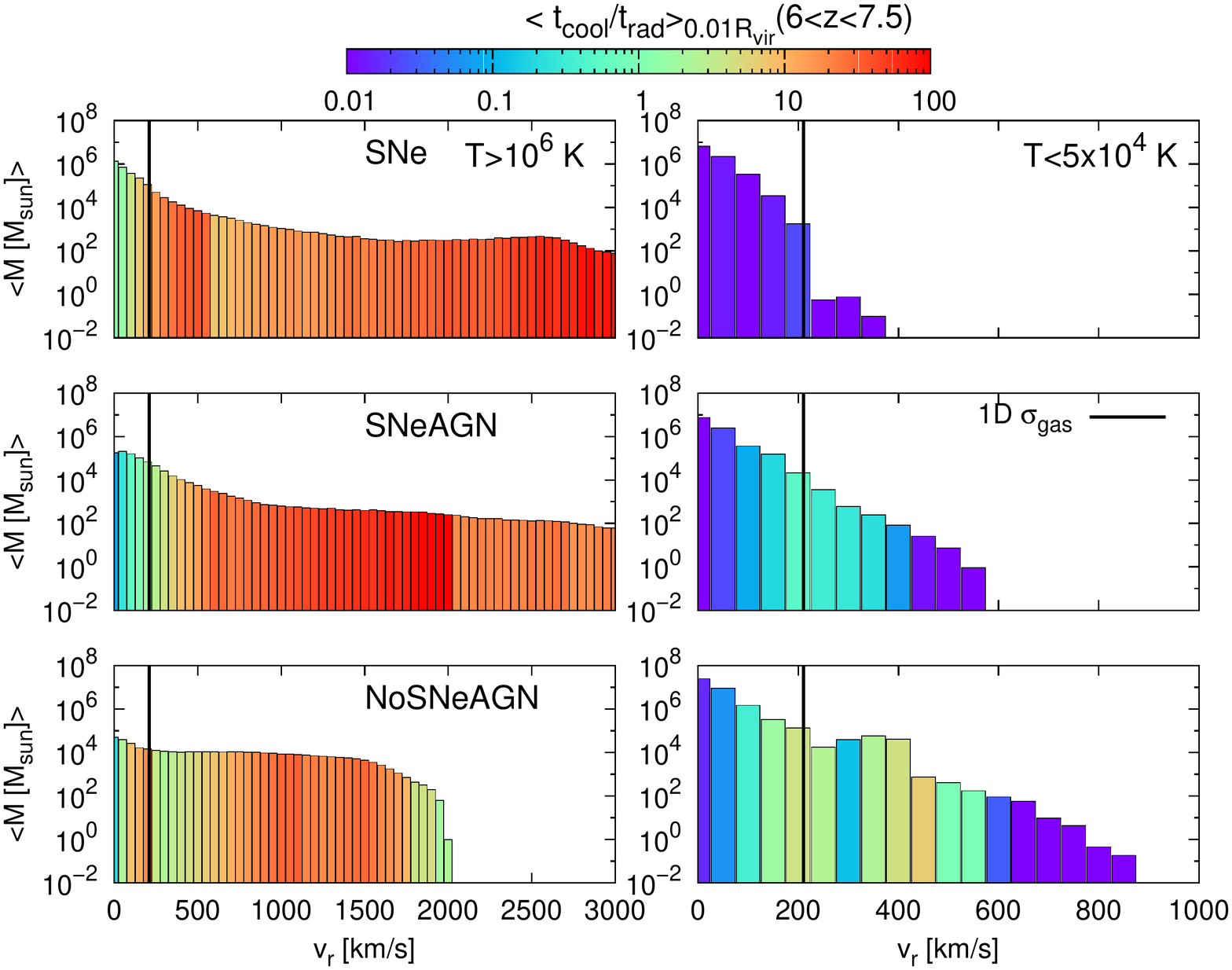}
\caption{Average mass in outflows radial velocity bins colored by 
the average cooling time 
to radial time ratio $\langle t_{\rm cool}/t_{\rm rad}\rangle$ for 
three of our simulations inside $0.01R_{\rm vir}$. The left column 
shows the hot $T>10^{6}$ K gas and the right column shows the 
cold $T>5\times 10^{4}$ K gas. The three panels are the average
PDFs for three different time intervals $\Delta t\approx 220$ 
Myr from $z\approx 15$ to $z=6$. The black solid line marks
the 1D gas velocity dispersion averaged over $\Delta t$ in the 
case of no feedback, NoSNe. See the text for discussion.}
\label{f9}
\end{figure}

Fig. \ref{f9} shows the average mass inside radial outflow velocity bins 
for gas inside $0.01R_{\rm vir}$ colored 
by the ratio $t_{\rm cool}/t_{\rm rad}$. The mass in each velocity bin 
is averaged over a time interval $\Delta t\approx 220$ Myr and they are
shown for three different redshift interval, namely $15 > z > 10$, $10 > z > 7.5$
and $7.5 > z > 6$. The cooling time scale is defined as $t_{\rm cool}=U/(n_{\rm H}\Lambda)$ 
where $U=3nk_{\rm B} T/2$ is the gas internal energy and $\Lambda$ is the gas 
cooling function, that is both a function of $T$ the gas temperature and $Z$ the 
gas metallicity. The radial time scale is defined as $t_{\rm rad}=r/v_{\rm r}$ 
where r is the radial coordinate and $v_{\rm r}$ is the gas outflow radial 
velocity with respect the BH motion. The origin of the system is set at the BH 
position. The left column of Fig. \ref{f9} shows the hot, $T>10^6$ K, gas and 
the right column shows the cold, $T<5\times 10^4$ K, gas. 
The black solid line marks the 1D velocity dispersion 
$\sigma_{\rm gas}$ of the NoSNe simulation averaged over $\Delta t$. This 
velocity dispersion gives us an idea about an outflow turbulent velocity 
triggered only by gravitational processes. In general, the feedback 
simulations presented here develop outflows with $v_{\rm r}\ga 2\sigma_{\rm gas}$.

When we look at the 
hot gas at high redshift, $z>10$, the NoSNeAGN run produces slower hot gas 
compared with all the SN feedback simulations. In this case the AGN activity 
can not accelerate the hot galactic gas above $\sim 400 \,\rm km\,s^{-1}$. In 
other words, without the SN heating the hot gas can not be accelerated above 
this velocity during the Eddington-limited growth phase of the BH. SN feedback 
can create a low density rarefied hot environment where energy injection 
from SNe (and AGN feedback for the SNeAGN run) can easily accelerate the gas.
In contrast, without SN feedback, the heated gas is surrounded by high density
gas and it is much more difficult to accelerate it at higher velocities. We
note that this fast hot gas is no more than a few percent of the total enclosed
gas mass. Note that this result differs quantitatively, but not qualitatively, 
from \citet{2015MNRAS.448L..30C} because they simulate a more massive halo of 
$3\times 10^{12} \,\rm M_\odot$ at $z=6$ powering a luminous quasar, while 
the halo in our simulations is of much lower mass 
$3\times 10^{10} \,\rm M_\odot$  and AGN feedback remains weak, and its 
effect limited. 

In the cold gas phase, clear differences exist between our 
experiments. Without AGN feedback, cold gas cannot be accelerated above 
$\sim 150-200 \,\rm km\,s^{-1}$, as shown also by 
\citet{2015MNRAS.448L..30C}. The inclusion of AGN feedback can accelerate 
cold gas to higher velocities, up to $\sim 300  \,\rm km\,s^{-1}$, for the 
case without SNe and up to $\sim 500  \,\rm km\,s^{-1}$, for the case with both 
SN and AGN feedback. This confirms, in a different regime, that combined effect 
of supernovae and AGN feedback boost each other and is what accelerates 
the cold gas: the AGN accelerates further the gas than has been already
accelerated by SNe. Finally, note that the large time scale ratios 
$t_{\rm cool}/t_{\rm rad}$ for the NoSNeAGN are caused by the lack of 
metal enrichment.

In the redshift interval $7.5<z<10$ the picture remains similar, but the now 
more massive BHs produce a more efficient AGN feedback, especially on the 
cold gas.
This efficiency increases with time, and at $6<z<7.5$ the NoSNeAGN simulation 
shows hot gas outflows up to $\sim 2000 \,\rm km\,s^{-1}$, but not beyond, in 
contrast with all the cases including SN feedback, which go as high as 
$\sim 3000 \,\rm km\,s^{-1}$. In the cold gas phase, SNe alone 
become completely ineffective at accelerating the cold gas, and it is only in 
the presence of an AGN that cold gas can be affected in the now massive galaxy 
(see the discussion in Dubois et al. 2015). Additionally, in the SNe only case, 
$\langle t_{\rm cool}/t_{\rm rad}\rangle \sim 0.01$ for the cold gas at 
$6<z<7.5$, while time ratios of order $\la 1$ are present in the simulations 
including AGN feedback. 

The cold gas in the no-Eddington limit simulation (not shown) can surpass
the $\sim 500  \,\rm km\,s^{-1}$ during super-Eddington accretion episodes 
at redshift above ten. Below $z=10$ the outflows can reach 
$\sim 1000 \,\rm km\,s^{-1}$ during super-Eddington episodes. 
Those episodes, however, are very short lived, and are followed by long quiescent periods, 
therefore the outflow velocity, averaged over hundreds of Myr, remains low.

Note that inside $0.01R_{\rm vir}$ the mass of hot gas is 
much lower than the mass of cold gas in our simulations. At  $z>10$ hot 
gas is not more than few percent of the total gas mass inside $1\%$ of the
virial radius, increasing to a few per cent at $7.5<z<10$. In the final
redshift range,  $6<z<7.5$, all the simulations including SN feedback reach
values in the range $\sim(10-25)\%$. The NoSNeAGN feedback remains at 
$\sim 1\%$ of hot gas and is the case with the lowest hot gas fraction also 
in the previous redshift intervals.

\subsubsection{BH and stellar mass evolution}
\begin{figure}
\centering
\includegraphics[width=1.0\columnwidth]{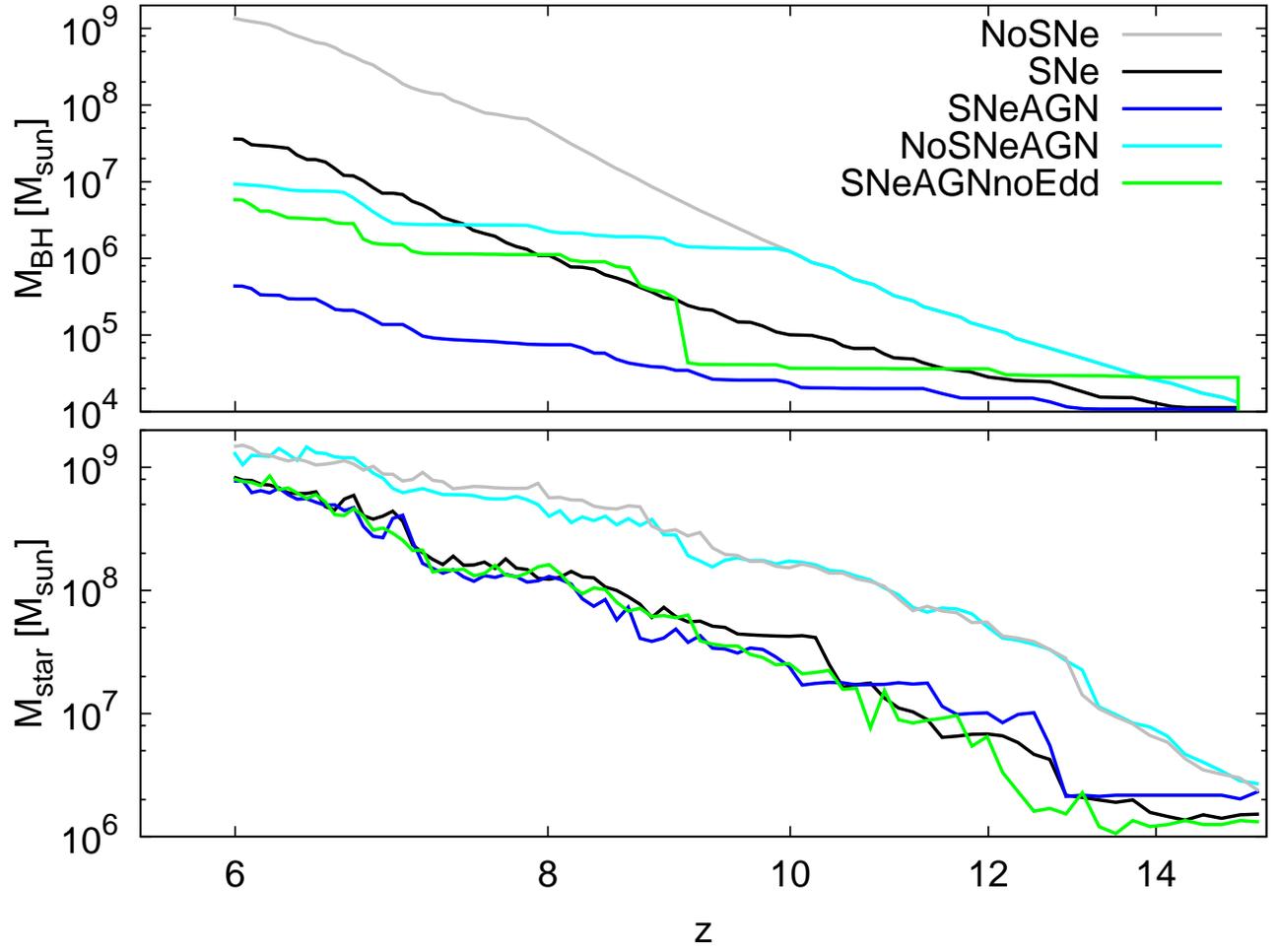}
\caption{BH mass (top) and stellar mass (bottom) as a function 
of redshift  for our four simulations. SNe in black, SNeAGN in 
blue, NoSNeAGN in cyan and SNeAGNnoEdd in green. The gray solid 
line in the top panel shows the BH mass in the no feedback 
simulation NoSNe of PE16. The NoSNeAGN run clearly forms
more stars, comparable to the NoSNe simulation in PE16.}
\label{f10}
\end{figure}

The discussion on the BH accretion rate and gas velocities presented in the 
previous sections now allows us to understand the BH and stellar mass evolution 
in this high-redshift galaxy. Fig. \ref{f10} shows the BH and 
galaxy stellar mass evolution as a function of redshift. The SNe simulation 
(black) reaches 
the highest BH mass at the end of the experiment (except for the run with no 
feedback at all described in PE16 and reported here for comparison). SN feedback 
can accelerate the gas because of pressure gradients associated to 
temperature differences of the order $\Delta T\sim 10^{5-6}$ K (depending on the 
gas density) in the galactic disc. 

As shown in the bottom left panel of Fig. \ref{f5}, due to the
high mass concentration around the BH (and its high escape velocity) 
the BH in the NoSNeAGN case has the fastest growth at high redshift amongst all 
our simulations, comparable with the no feedback case of PE16. As shown in table 
\ref{table}, the BH grows at the Eddington rate at $z>10$, until it reaches 
$M_{\rm BH}\approx 10^6$ M$_\odot$.  At this stage  AGN feedback is capable 
of accelerating the surrounding  gas beyond the central escape velocity, 
reducing dramatically the BH  growth rate. The BH ends the simulation with a 
mass of $M_{\rm BH}\approx 9\times 10^6$ M$_\odot$. 

At the beginning of the evolution, the SNeAGNnoEdd BH shows a large mass 
jump, increasing almost by three times its initial value due to short bursts of 
super-Eddington accretion accompanied by strong feedback. Afterward, as
discussed 
in the previous sections, the central region is devoid of gas and BH growth is 
stunted until a merger at $z\sim 9.5$ that replenishes the gas supply and the 
BH experiences a high super-Eddington accretion ($f_{\rm EDD}\la 10$) episode 
increasing its mass by almost one order of magnitude. By this time, however, 
AGN feedback is not as disruptive anymore in the denser and more massive nucleus 
and the BH continues its evolution 
with irregular mass accretion rate and with more super-Eddington episodes.

The SNeAGN feedback has the slowest mass growth and then it reaches the 
lowest mass at the end of the simulation ($M_{\rm BH}\approx 4\times 10^5$ 
M$_\odot$). In 
this case, beside the SNe outflows,  AGN activity can 
create outflows associated to temperature gradients of $\Delta T\ga 10^{7}$ K, 
much larger than the ones produced by SNe. This result shows that the combined 
effect of SNe and AGN feedback works together to quench efficiently 
the BH growth.

\begin{figure}
\centering
\includegraphics[width=15cm,height=15cm]{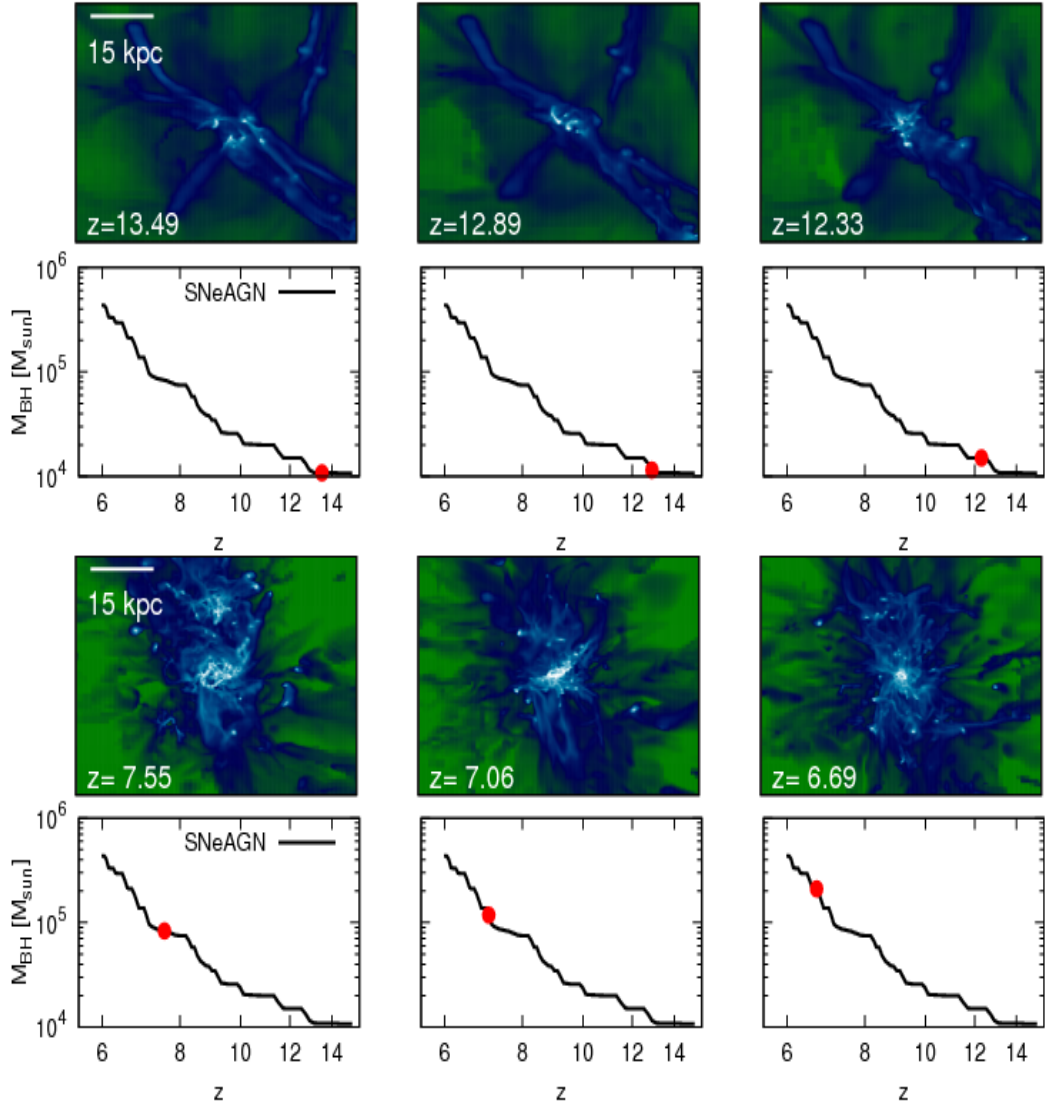}
\caption{Gas maps (from top to bottom, the first and third rows) and 
BH mass evolution (from top to bottom, the second and fourth rows) as 
a function of redshift for the SNeAGN run. The figure shows the 
BH mass evolution at different redshifts marked with red dots. 
These are two examples of the connection between merger events 
and rapid BH growth in our simulation.}
\label{f11}
\end{figure}

At ${\rm z\approx 7}$ all the  simulations including AGN feedback 
(SNeAGN, NoSNeAGN and SNeAGNnoEdd) show a sudden change in the BH mass. 
At this redshift the system experiences a $1:3$ merger, bringing fresh high 
density gas to the central galactic region to fuel the BH. 
Fig.  \ref{f11} shows two episodes of rapid BH mass growth for the SNeAGN 
simulation, one at ${\rm z\approx 13}$ and other one at ${\rm z\approx 7}$ 
(mentioned above). In this two examples it is possible to associate a merger 
event with a jump in the BH mass evolution \citep[see][]{Dubois+2015}. 

Regarding the stellar mass evolution, the simulation with no SNe and only 
AGN feedback, NoSNeAGN, 
produces the galaxy with the highest stellar mass at the end of the simulation,
comparable but lower to the NoSNe run in PE16. The final 
stellar mass is $M_{\star}\approx 1.5\times10^9$ M$_\odot$. As expected, the 
lack of fast outflows allows high gas concentration, and therefore an efficient 
star formation throughout the simulation. In contrast, all the cases including 
SN feedback finish the simulation with a lower stellar mass, of the order of 
$M_{\star}\approx 8\times 10^8$ M$_\odot$, showing that the BH mass is not
large enough in these galaxies to quench star formation. 

All the SN feedback simulations have a similar stellar masses at 
$z< 8$, $M_{\star}> 10^8$ M$_\odot$. Above this redshift it is possible to see 
that the stellar masses differ. In other words, it seems that below 
$M_{\star}\sim10^8$ M$_\odot$ SNe and AGN feedback produce a larger scatter in 
the stellar mass content at similar DM halo mass.   

\section{Discussion and Conclusions}
\label{conclusion}

We have run cosmological zoom-in simulations of high redshift galaxies in order 
to study the effect of SN and AGN feedback on the mass transport in these
objects and their consequences on the central BH growth. 

As in PE16 we find that the mass accretion rate beyond the virial radius 
is of the order of $\sim 10$ M$_\odot\,\rm yr^{-1}$. These high mass 
accretion rates are associated to material in almost free-fall going onto 
the central DM halo region. In these systems,  feedback from SNe and 
AGN is able to suppress the low density mass accretion at the galactic edge.
Material with densities associated with cooling flows, i.e. 
$n\la 0.1-1$ cm$^{-3}$ can not penetrate inside the central $\sim$ kpc. 
Only high density $n\ga 10^2$ cm$^{-3}$ gas is able to reach the inner 
$\sim$ few 100 pc. The simulations with only SNe, and the simulation with 
SNe plus AGN feedback are those for which the suppression is the highest, 
in contrast to the AGN only experiment where gas can reach the 
galactic center more easily.

In these turbulent galactic environments, torques acting on the gas have 
two sources: gravity force associated to an inhomogeneous mass distribution 
and pressure gradients associated to circum-galactic shocks driven by 
cosmic infall, or SN and AGN feedback. Such torques are required to 
re-distribute the gas AM from the external edge of the galaxy to 
the central regions, at ${\sim}$ pc scales, and trigger a radial mass accretion rate of 
$\dot{M}\sim 1$ M$_\odot\,\rm yr^{-1}$ .

All the SN feedback simulations produce a lower 
$\langle f_{\rm EDD}\rangle$ at $z>10$ compared with their 
$\langle f_{\rm EDD}\rangle$ at $z<10$, showing that  SN feedback can 
quench efficiently  BH growth in small galaxies at high redshift
until a critical mass is reached \citep{Dubois+2015}. Furthermore, it is possible to see that after 
merger events, the BH has a significant growth. During such events a large amount 
of gas reaches the BH, feeding it with fresh gas. 

Our four simulations show very different BH mass accretion histories depending
on the physical ingredients we included. Although SN feedback alone is 
able to alter dramatically the BH accretion rate 
\citep[][and PE16]{Dubois+2015,2016arXiv160509394H},
the SN and AGN feedback in tandem are the most efficient to quench the 
BH growth. Due to the high gas density around the BH at high redshift, 
the simulation
without SN feedback can grow near the Eddington limit until it 
reaches the $M_{\rm BH}\sim 10^6\rm M_\odot$, and self-regulates by its 
AGN activity.

Regarding stellar mass growth,  AGN feedback alone is also unable to 
regulate SF,
although the BH-stellar mass ratio is large, in fact for the noSNeAGN 
case, the ratio is well
above $10^{-3}$ at almost all times. SN feedback, in these small 
galaxies, seems to be much more effective at affecting SF. 

Our simulations show that most of the central gas inside $1\%$ of 
the halo virial radius is cold ($T<5\times 10^4$ K). The hot ($T>10^6$ K) gas 
is no more than a few percent above redshift $z\ga7.5$.
Below this redshift, all the simulations including SN feedback increase 
the amount of 
hot gas reaching fractions $\la (10-25)\%$. In contrast, the no SN 
feedback simulation keeps its hot gas fraction very low $\la 1\%$ over 
its entire evolution.

The simulation without SN feedback shows that it is not possible to accelerate 
the hot gas beyond $2000 \rm\, km\,s^{-1}$. In contrast all the SN 
feedback simulations 
can easily surpass this value, showing that SN heating creates a low density
rarefied gas where the SN and AGN feedback accelerate the gas at
velocities as high as $3000 \rm\, km\,s^{-1}$.

The cold gas phase in the innermost galactic region of the SN 
feedback simulation can not be accelerated beyond 
$\sim 150-200 \rm\, km\,s^{-1}$ 
\citep{2015MNRAS.448L..30C} whereas with AGN plus SN feedback the cold gas can
reaches up to $\sim 500 \rm\, km\,s^{-1}$, showing that the combined effect 
of SN and 
AGN heating creates the strongest outflows of cold gas and consequently 
is the most efficient combination to quench the BH and galaxy growth at 
high redshift.  

We note that our results should depend on the BH feedback prescription.
In particular a more realistic approach based on a quasar and radio-jet-like
mode as in \citet{Dubois+2012agn} would produce a lower effect on the 
surrounding gas increasing the BH growing rate and reducing the gas outflows
velocity. A simulation including a jet like BH feedback will be presented 
in an up-coming paper.

\acknowledgments
Powered@NLHPC: This research was partially supported by the supercomputing
infrastructure of the NLHPC (ECM-02). The Geryon cluster at the Centro de
AstroIngenieria UC was extensively used for the analysis calculations performed
in this paper. J.P. acknowledges the support from proyecto anillo de ciencia y
tecnologia ACT1101. A.E. acknowledges partial support from the Center of
Excellence in Astrophysics and Associated Technologies (PFB06), FONDECYT
Regular Grant 1130458. MV acknowledges funding from the European Research 
Council under the European Community's Seventh Framework Programme 
(FP7/2007-2013 Grant Agreement no.\ 614199, project ``BLACK''). Part of this
work has been done within the Labex ILP (reference ANR-10-LABX-63) part of 
the Idex SUPER, and received financial state aid managed by the Agence
Nationale de la Recherche, as part of the programme Investissements 
d'avenir under the reference ANR-11-IDEX-0004-02.

\begin{table}
\begin{center}
\caption{Normalized BH accretion rates and BH masses at $z=10$ and $z=6$.}
\begin{tabular}{l c c c c c c}
\hline\hline
            &                        &                        &           &                   & \\
Simulation  & $\langle f_{\rm Edd}\rangle$ & $\langle f_{\rm Edd}\rangle$ & $\langle f_{\rm Edd}\rangle$ & $M_{\rm BH}$ (z=6) & $M_{\rm BH} (z=10)$\\
            &                        & $z>10$                 & $z<10$    & $\rm M_\odot$              & $\rm M_\odot$   \\
\hline 
            &                        &                        &           &                   & \\
SNe         & 0.54                   & 0.48                   & 0.57      & $3.6\times10^7$   & $1.0\times10^5$\\
SNeAGN      & 0.24                   & 0.14                   & 0.30      & $4.3\times10^5$   & $2.0\times10^4$\\    
NoSNeAGN    & 0.49                   & 0.99                   & 0.21      & $9.3\times10^6$   & $1.1\times10^6$\\  
SNeAGNnoEdd & 0.32                   & 0.06                   & 0.50      & $5.8\times10^6$   & $3.7\times10^4$\\
\hline
\end{tabular}
\label{table}
\end{center}
\end{table}

\end{document}